\documentclass[pre,aps,preprint,superscriptaddress]{revtex4-1}
\usepackage{graphicx}
\usepackage{amsmath}
\usepackage{epstopdf}
\begin{document}
\title{Analysis of Geometric Phases in Particle Diffusion Systems: Insights from Non-Hermitian Heat Transfer}
\author{Jinrong Liu}
\author{Liujun Xu}
\author{Gaole Dai} 
\author{Gang Wang}
\begin{abstract}
Geometric phases in particle diffusion systems, an intriguing aspect enlightened from thermal systems, offer a different understanding beyond traditional Brownian motion and Fick's laws. This concept introduces a phase factor with significant implications for particle behavior, central to which is the non-Hermitian nature of the Hamiltonian in the diffusion system. A unique structure composed of two rings moving in opposite directions and a stationary intermediate layer plays multifunctional roles in controlling particle diffusion. 
For the real-world application, a bilayer particle-diffusion cloak is demonstrated, showcasing the extensive control and adaptability achievable through mastering geometric phases.
This system has potential applications in industries like healthcare and environmental management, thus expanding the understanding of the geometric phase and offering insights for the design of particle-diffusion metamaterials.

\end{abstract}
\maketitle
\section{Introduction}
Particle diffusion is a fundamental concept that describes the random motion of particles suspended in a fluid, resulting from their collision with the fast atoms or molecules in the gas or liquid. The traditional understanding of particle diffusion is based on the laws of Brownian motion and Fick's laws of diffusion \cite{ljrFaucheuxPRE94,ljrVanMilligenEJP05,ljrNguyenJSP19,ljrShankarNE20,ljrKayJPA23}. However, real systems are often heterogeneous, and the traditional laws might not hold universally. The uniformity or non-uniformity of particle distribution can significantly alter the permeable \cite{ljrMalgarettiEPL23,ljrSposiniCP22}, mechanical \cite{ljrFangCPL07,ljrTianEPL07}, thermal \cite{ljrYangJAP07,ljrHuangSSC00} and electrical \cite{ljrXiaoJPCB2008,ljrGaoAPL2008,ljrZhangAPL08,ljrJianJRCB08,ljrXuJMR08,ljrFanJPCB06, ljrHuangPRE05a, ljrHuangPRE04b, ljrHuangJPCM04, ljrHuangCP04, ljrHuangJPCM02} properties of composite materials. The interaction between particles and the fluid medium can lead to various emergent phenomena, such as altered electrical or optical properties \cite{ljrFanJAP2008,ljrGaoJPCC07,ljrWangJPCB06,ljrXiaoJPCB2008,ljrGaoAPL2008,ljrZhangAPL08,ljrJianJRCB08,ljrXuJMR08,ljrYePA081,ljrChenJPA07}.
Moreover, Fick's law has been shown to be form-invariant under coordinate transformations, leading to the emergence of particle-diffusion metamaterials. This novel field offers potential for manipulating diffusion in new ways \cite{ljrGuenneauJRSI13,ljrGuenneauAIPA15,ljrRestrepoSciRep16,ljrRestrepoJAP16,ljrRestrepoAPL17,ljrRestrepoJPD17}.


Beyond the conventional understanding of particle diffusion, there exists an intriguing aspect called the geometric phase \cite{ljrXuPRE20} enlightened from thermal system \cite{ljrLeiMTP23,ljrZhuangIJMSD23,ljrZhouEPL23}. This concept diverges from traditional diffusion models \cite{ljrXuCPL20,ljrYangPRAP20,ljrWangiScience20,ljrXuEPL20}, introducing a phase factor with notable implications for particle behavior. Central to this concept is the non-Hermitian Hamiltonian $H$ in the diffusion system. This Hamiltonian, which has been effectively characterized in two countermoving media \cite{ljrLiScience19}, offers a fresh view on diffusion. It underscores the significance of geometric and topological factors in the diffusion process \cite{ljrOzdemirNM19,ljrWangNature22,ljrEsakiPRB11,ljrYaoPRL18,ljrTakataPRL18}.  What's fascinating about its non-Hermitian nature is its potential to reveal diffusion behaviors not encompassed by traditional Hermitian quantum mechanics \cite{ljrLiScience19,ljrXuPRL21,ljrXuNP22,ljrXuPNAS22}. 

In particle diffusion, the geometric phase introduces unexpected behaviors and effects not anticipated by traditional theories. One such effect is a deviation from standard diffusion patterns, manifesting as anomalous diffusion or new modes of diffusion \cite{ljrXuPRE20}. These deviations can significantly impact fields from material science to biology, where precise control and comprehension of diffusion are vital.

\section{Theory of Non-Hermitian diffusion systems}

\begin{figure}[ht]
    \includegraphics[width=\linewidth]{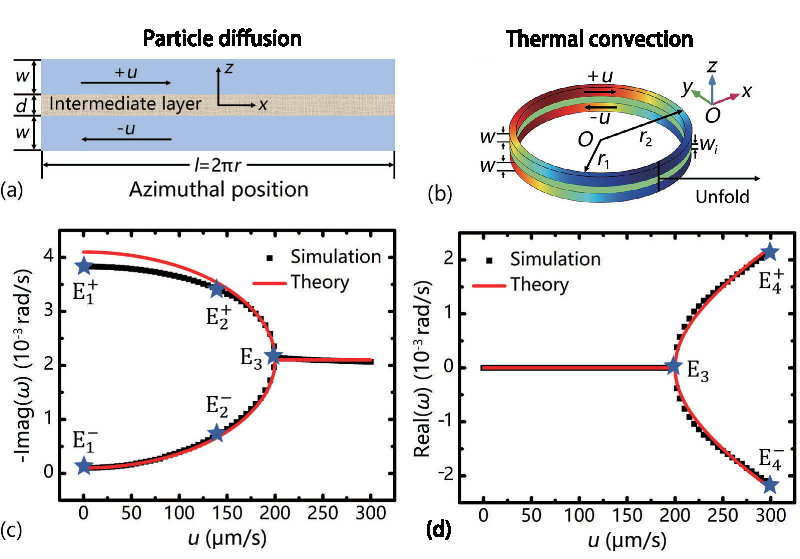}
    \caption{This figure provides a comprehensive view of the topological diffusion system. Part (a) showcases a simplified two-dimensional particle diffusion model with specific parameters while Part (b) shows the thermal convection system. In particle diffusion model, parts (c) and (d) plot the negative imaginary and real parts of the eigenvalue $ \omega $ as a function of velocity $ u $. Solid lines and block squares represent analytical and simulation results, respectively. Adapted from \cite{ljrXuPRE20} and \cite{ljrXuIJHMT21} with author's permission.}
\end{figure}

Before diving deeper into the geometric phase's role in particle diffusion systems, focus should be turned to the intricate configurations that induce non-Hermiticity. A pivotal study in thermal domain has outlined a fundamental architecture comprising two dynamically moving rings and a stationary intermediary layer \cite{ljrLiScience19}, which has been further developed in certain thermal systems \cite{ljrXuPRL21,ljrXuNP22}. Fig. 1 vividly illustrates the foundational components of thermal structure \cite{ljrXuIJHMT21} and particle diffusion \cite{ljrXuPRE20}. In both systems, the rings move with equal velocities but opposite directions. The intermediary layer acts as a medium for interchange between these two dynamic rings, where the interchange is the particle exchange in Fig. 1a and heat transfer in Fig. 1b. To formulate this topological process mathematically, a non-Hermitian Hamiltonian $H$ is employed:

\begin{equation}
H = \begin{pmatrix}
-i(k^2D + h) + ku & ih \\
ih & -i(k^2D + h) - ku
\end{pmatrix}
\end{equation}

In this expression, $k$ signifies the wave number, $D$ denotes the diffusivity of the two dynamic rings, $u$ represents the velocity, and $i = \sqrt{-1}$ is the imaginary unit. The term $h = \frac{D_m}{w \cdot d}$ encapsulates the rate of particle interchange between the two dynamic rings, where $D_m$ is the diffusivity of the intermediary layer, and $w$ and $d$ are the thicknesses of the dynamic rings and the intermediary layer, respectively \cite{ljrXuPRE20}.


The evolution of eigenstates in a particle diffusion system with geometric phase is a subject of considerable interest. The eigenstates are not static but evolve over time, especially when the system involves non-Hermitian dynamics. The initial and final states of the system can be significantly different, depending on the parameters such as the velocity of the moving rings and the properties of the intermediate layer. The eigenvalues and eigenstates of this Hamiltonian in Eqs. 1 are given by:
\begin{equation}
\omega = -i \left[ (k^2 D + h) \pm \sqrt{h^2 - k^2 u^2} \right]
\end{equation}
\begin{equation}
\psi_+ = 
\begin{cases} 
[1, e^{i(\pi-\theta)}]^T, & \text{if } u < u_{\text{EP}} \\
[e^{-\phi}, e^{i\pi/2-2\phi}]^T, & \text{if } u > u_{\text{EP}}
\end{cases}\quad
\psi_- = 
\begin{cases} 
[1, e^{i\theta}]^T, & \text{if } u < u_{\text{EP}} \\
[e^{-\phi}, e^{i\pi/2}]^T, & \text{if } u > u_{\text{EP}}
\end{cases}    
\end{equation}
where $\varphi = \cosh^{-1}(ku/h)$, $u_{\text{EP}}=h/k$. This unified expression provides a comprehensive view of how the eigenstates evolve depending on the velocity parameter \( u \), and how they are influenced by the exceptional point \( u_{\text{EP}} \) in Fig. 1. The evolution of these eigenstates is influenced by the geometric phase, which can be expressed as:

\begin{equation}
\gamma_\pm = i \int \frac{\langle \psi_\pm(u), d\psi_\pm(u) \rangle}{\langle \psi_\pm(u), \psi_\pm(u) \rangle}
\end{equation}

This geometric phase accumulates over time and plays a crucial role in the system's dynamics. For instance, when the system evolves around an exceptional point, the geometric phase takes on values of $\pi$ or $-\pi$, depending on the direction of the closed loop. If the evolution route does not contain the exceptional point, the integral in a closed loop is naturally equal to zero.

In summary, the evolution of eigenstates and the accumulation of geometric phase are pivotal in understanding the complex dynamics of particle diffusion systems with geometric phase. These factors are not only theoretically intriguing but also have practical implications in designing systems for effective manipulation of particle diffusion.

\section{Numerical Simulations of Eigenstate Evolution and Geometric Phase}

\begin{figure}[ht]
    \includegraphics[width=0.6\linewidth]{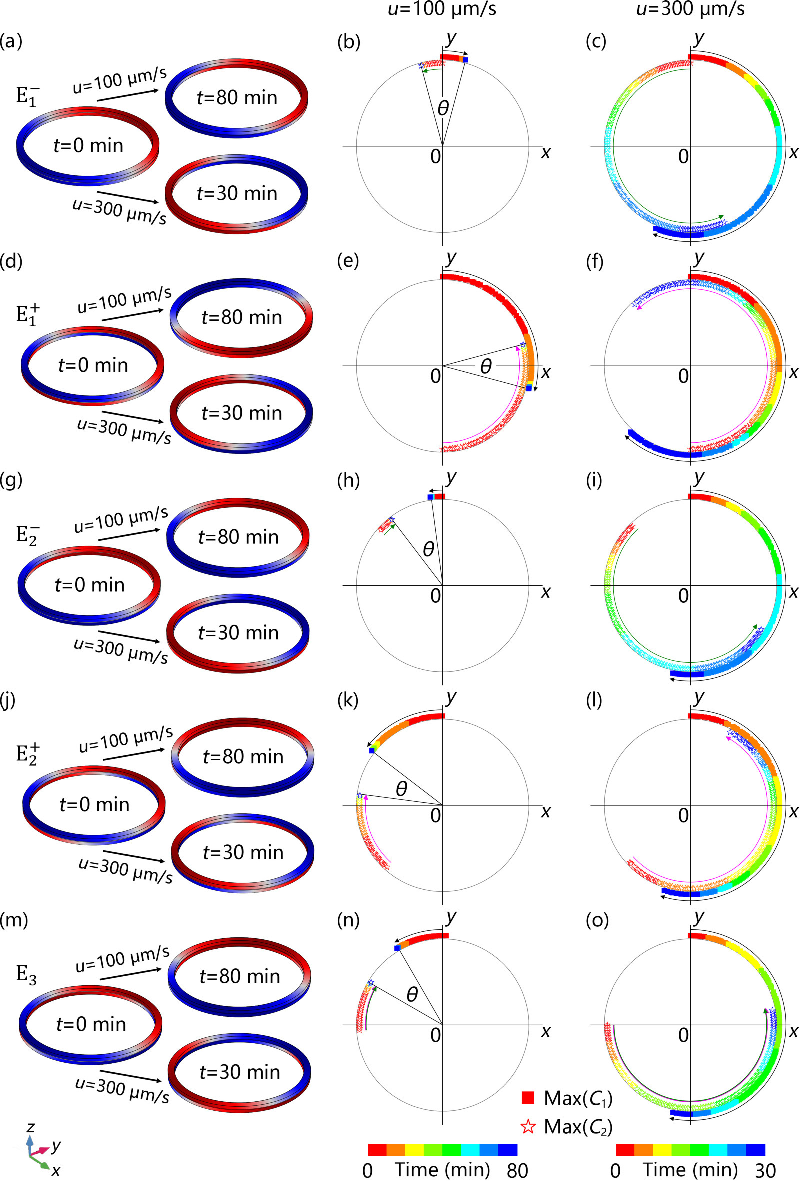}
    \caption{This figure presents the initial and final states of the system at two different velocities. It also depicts the trajectories of the maximum concentrations $ \text{Max}(C1) $ and $ \text{Max}(C2) $ along the interior edges of the two moving rings. The parameters used for the simulations are $ w = 0.5 \, \text{cm} $, $ d = 0.1 \, \text{cm} $, $ r = 10 \, \text{cm} $, $ r' = 11 \, \text{cm} $, $ D = 10^{-6} \, \text{m}^2/\text{s} $, and $ D_m = 10^{-8} \, \text{m}^2/\text{s} $. Adapted from \cite{ljrXuPRE20} with author's permission.}
\end{figure}

Relying on the theoretical principles of eigenstate evolution and geometric phase, numerical analyses are executed utilizing COMSOL Multiphysics. 
A striking resemblance emerges between the dynamics within these particle diffusion systems and those in thermal convection-conduction systems. This similarity is grounded in the fact that both thermal convection and particle diffusion are manifestations of diffusion metamaterial systems, governed by a similar class of partial differential equations. Drawing from this observation, a recent thermal study \cite{ljrXuIJHMT21}, substituting the diffusivity $D$ with the thermal coefficient ${\kappa}/{\rho c}$, the topological exceptional points (EPs) and the evolution pathways of states traverse the same topological branch. Despite the parallels, it's vital to discern the distinctive nuances that differentiate each system, thereby enriching our understanding of diffusion phenomena across various platforms.

Utilizing the finite element methods available in COMSOL Multiphysics, the simulations offer a granular perspective on the behavior of the geometric phase. Distinct velocity paths are selected for the simulations; some are designed to circumvent the exceptional point (EP), while others intentionally intersect with it. The simulations commence with an initial velocity of $100\ \mu m/s$ and an initial state corresponding to the eigenvalue $\omega^-$, characterized by a phase difference of $\pi/6$. In scenarios where the system's trajectory avoids the EP, the eigenvalue remains purely imaginary, thereby confirming the absence of any accrued phase difference. This specific trajectory ensures that the system reverts to its original concentration profile, as depicted in Fig. 2.

Conversely, when the system's trajectory encompasses the EP, a dramatic shift in the dynamics is observed. The EP serves as a critical nexus where the eigenvalues coalesce, resulting in a Hamiltonian that is non-diagonalizable. As the system traverses this point, real components of the eigenvalues emerge, signifying the introduction of an additional phase difference, thereby enriching our understanding of the geometric phase's influence on the system's behavior.

\begin{figure}[ht]
    \includegraphics[width=0.6\linewidth]{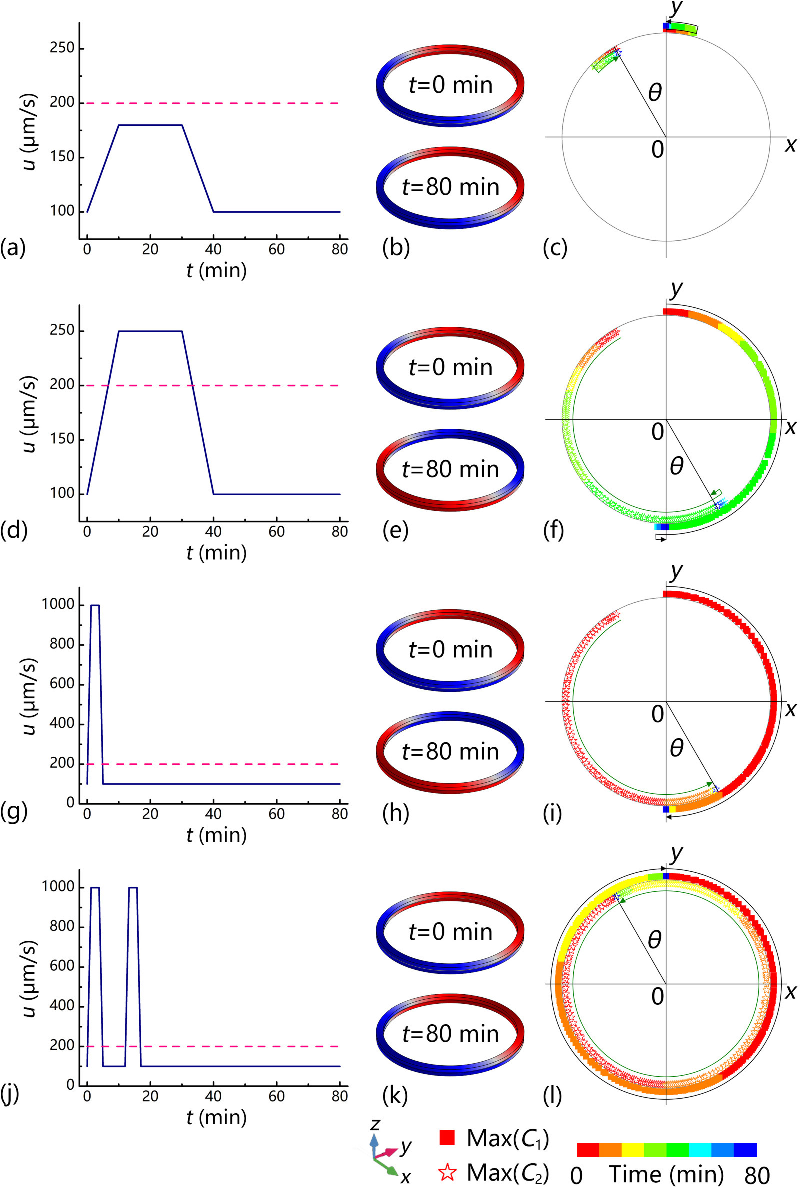}
    \caption{The figure is divided into three columns. The left column outlines the time-varying paths of velocity. The middle column displays the initial and final states of the system, while the right column illustrates the trajectories of $ \text{Max}(C1) $ and $ \text{Max}(C2) $ along the interior edges of the channels. The parameters are consistent with those used in Fig. 2. Adapted from \cite{ljrXuPRE20} with author's permission.}
\end{figure}

To deepen the exploration of the geometric phase, the simulations also scrutinize the transition from non-eigenstates to eigenstates. Initial states are selected from a set of five distinct eigenstates, each defined in Cartesian coordinates. Velocities for these simulations are set at $100 \, \mu m/s$ and $300 \, \mu m/s$, and the directions of velocity are clockwise for the upper ring and anticlockwise for the lower ring. The theoretical phase differences at $u = 100 \, \mu m/s$ are $\pi/6$ for $\omega^-$ and $5\pi/6$ for $\omega^+$. Intriguingly, these initial states, which are not eigenstates at the outset, evolve to align with the eigenstates. Ultimately, all five initial states converge to a singular final state, corresponding to the eigenvalue $\omega^-$. This convergence is ascribed to the non-orthogonality of the eigenstates across different branches, as illustrated in Fig. 3.

In summary, the numerical simulations executed in COMSOL Multiphysics not only substantiate the theoretical framework but also furnish a nuanced understanding of the role played by the geometric phase in particle diffusion systems. These insights are pivotal for guiding future experimental validations and practical applications.

\section{Bilayer Particle-Diffusion Cloak: Design and Applications}

Expanding beyond the theoretical realm of eigenstate evolution and geometric phase, the focus shifts to the cutting-edge application of the bilayer particle-diffusion cloak, a concept also viable in thermal convection systems due to analogous principles in play. This sophisticated design involves a dynamic interplay between two concentric moving rings and a static intermediate layer, leveraging the remarkable adaptability inherent in both particle diffusion and heat transfer scenarios. 
In thermal systems, the intermediate layer, while stationary, experiences an effective conductivity enhancement due to the enhanced surface flux, making the cloak applicable even in highly conductive backgrounds. This enhancement is facilitated by the competition between conduction and convection in the intermediate layer, significantly influenced by the motion of the two rings, as detailed in thermal studies. Similarly, in particle diffusion systems, the cloak's effectiveness hinges on fine-tuning the individual rotational speeds of the rings, allowing for a robust cloak in different diffusivity $D$ backgrounds, even within rotational frames of reference.

The architectural blueprint of this cloak is a two-tiered strategy. The first tier involves the deployment of a zero-diffusivity layer to encapsulate the object, thereby nullifying its diffusion influence on the surrounding environment. The second tier incorporates an additional layer consisting of two moving rings. These rings are meticulously engineered to counterbalance the diffusion effects of the isolated region. By judiciously modulating the velocities of these rings, which are antiparallel but equal in magnitude, a state of particle-diffusion invisibility can be realized.
\begin{figure}[ht]
    \includegraphics[width=0.7\linewidth]{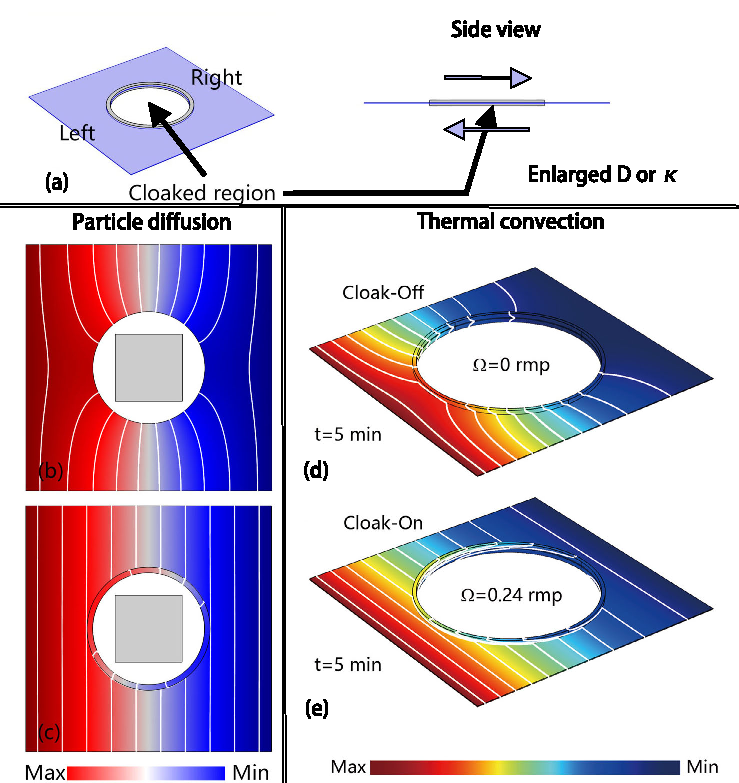}
    \caption{This figure demonstrates the cloaking effect in a linear concentration field and thermal field. The cloaking effect is achieved by enhancing the diffusivity $D$ or thermal conductivity $\kappa$ of the intermediate layer. Adapted from \cite{ljrXuPRE20} and \cite{ljrXuIJHMT21} with author's permission.}
\end{figure}
To quantify the cloak's effectiveness, the enhanced diffusivity \( D' \) is given by the formula:

\[
D' = D \left( 1 + \frac{{v^2 \tau^2}}{4} \right),
\]
where \( D \) is the original diffusivity, \( v \) is the velocity of the moving rings, and \( \tau \) is the relaxation time. Numerical simulations, represented in Fig. 4, corroborate the cloak's efficacy under both linear and nonlinear diffusion regimes. Specifically, the simulations indicate that a velocity setting of 37 $\mu$m/s for the moving rings yields an enhanced diffusivity that meets the criteria for effective cloaking. This results in an undistorted concentration profile in the background, thereby affirming the cloak's operational success.

A notable strength of this bilayer cloak lies in its adaptability and straightforward implementation. Standing in stark contrast to other cloaking techniques that require complex transformation optics or detailed multi-layered infrastructures, this innovative approach is grounded in consistent parameters and simple geometries. Such foundational simplicity underscores its practicality, enhancing its suitability for a plethora of real-world scenarios. A further testament to its user-friendly design is the cloak's operational versatility. The cloaking function can be effortlessly engaged or disengaged by altering the speed of the moving rings, demonstrating flexibility in various application settings. Similarly in thermal dynamics, this cloak-on and cloak-off feature can be achieved. The thermal cloak can be efficiently activated or deactivated in transient states, providing a distinct advantage over particle diffusion systems, which are constrained to steady-state operations.

Embarking on the quest to make particle diffusion invisible constitutes a significant breakthrough, heralding a spectrum of applications that permeate sectors as diverse as chemical engineering and biological systems. More than just a mechanism for concealment, this technology is strategically devised to navigate entities through a range of environments, introducing groundbreaking possibilities in areas like pollution abatement, precision-targeted healthcare delivery, and discreet environmental monitoring. This advanced level of control and finesse positions the bilayer particle-diffusion cloak as a pioneering development, reshaping traditional approaches to managing diffusion phenomena. Consequently, it ushers in a new benchmark for operational excellence across interdisciplinary scientific and industrial landscapes.

\section{conclusion}
The field of geometric phases in particle diffusion is not merely an academic endeavor but a multidisciplinary area with far-reaching implications. The paper began by laying the theoretical groundwork, elucidating the basic principles and significance of geometric phase in diffusion systems. It then delved into the intricacies of designing structures that can effectively manipulate particle diffusion, followed by a rigorous examination of eigenstate evolution and the role of geometric phase. The concept of a bilayer particle-diffusion cloak was introduced, offering a novel approach to controlling diffusion processes. Finally, the paper culminated in discussing the real-world industrial applications, which are substantiated by methodological advancements and empirical research.

The interplay between theory, simulation, and practical application serves as a testament to the field's dynamism and its potential for groundbreaking innovations. While the paper has provided a comprehensive overview, it is important to acknowledge that the field is continually evolving. New methodologies are being developed, novel applications are being discovered, and further research is being conducted to validate and refine existing theories. 

In conclusion, the exploration of geometric phases in particle diffusion systems is a burgeoning field that promises to revolutionize both scientific understanding and industrial practices. The theoretical frameworks, simulation models, and industrial applications discussed in this paper are but a snapshot of what the field has to offer. As research progresses, it is anticipated that the coming years will witness even more remarkable advancements, further solidifying the importance of this area in both academic and industrial landscapes.

\end{document}